\documentclass[twocolumn,showpacs,floats,floatfix,superscriptaddress,aps,pra]{revtex4-1}
\usepackage{amsfonts}
\usepackage{amssymb}
\usepackage{amsmath}
\usepackage{calc}
\usepackage{graphicx}
\usepackage{bm}

\usepackage[normalem]{ulem}
\usepackage{amsmath,amssymb}
\usepackage{multirow}
\usepackage{xcolor,soul}
\usepackage{srcltx}
\usepackage{hyperref,graphicx}

\def\be{ \begin{equation}}
\def\ee{ \end{equation}}
\def\bea{ \begin{eqnarray}}
\def\eea{ \end{eqnarray}}
\def\bse{ \begin{subequations}}
\def\ese{ \end{subequations}}
\def\bc{ \begin{center}}
\def\ec{ \end{center}}

\begin{document}

\author{Stefano Longhi$^{*}$} 
\affiliation{Dipartimento di Fisica, Politecnico di Milano and Istituto di Fotonica e Nanotecnologie del Consiglio Nazionale delle Ricerche, Piazza L. da Vinci 32, I-20133 Milano, Italy}
\email{stefano.longhi@polimi.it}

\title{Tight-binding lattices with an oscillating imaginary gauge field}
  \normalsize

%\date{.}

%
\bigskip
\begin{abstract}
\noindent  
We consider non-Hermitian dynamics of a quantum particle hopping on a one-dimensional tight-binding lattice made of $N$ sites with asymmetric hopping rates induced by a time-periodic oscillating imaginary gauge field. A deeply different behavior is found depending on the lattice topology. While in a linear chain (open boundary conditions) an oscillating field can lead to a complex quasi energy spectrum via a multiple parametric resonance, in a ring topology (Born-von Karman periodic boundary conditions) an entirely real quasi energy spectrum can be found and the dynamics is pseudo-Hermitian. In the large $N$ limit, parametric instability and pseudo-Hermitian dynamics in the two different lattice topologies are physically explained on the basis of a simple picture of wave packet propagation. 
\end{abstract}

\pacs{03.65.-w,  72.20.Ee, 72.15.Rn,  73.43.-f, 71.30.+h }

% 03.65.-w Quantum mechanics
% 72.20.Ee    Mobility edges; hopping transport
%42.25.Bs, 42.82.Et, 11.30.Er 
% 11.30.Er 	Charge conjugation, parity, time reversal, and other discrete symmetries
% Supersymmetry,
% 72.15.Rn: Localization effects
%71.30.+h Metal-insulator transitions and other electronic transitions
%  73.43.-f Quantum Hall effects

\maketitle

\section{Introduction}
 Non-Hermitian models have attracted since many years a
considerable attention in different areas of physics with
 applications in a variety of fields, including quantum mechanics of open systems \cite{r1,r2}, parity-time ($\mathcal{PT}$) symmetric quantum mechanics and quantum field theories \cite{r3}, atom optics \cite{r4},  hydrodynamics \cite{r5}, superconductors \cite{r6},  biological \cite{r6bis} and optical \cite{r7,r8,r9} systems to mention a few.\par Among various non-Hermitian quantum models, great attention has been devoted to the study of the hopping dynamics of a quantum particle in a lattice in the presence of an 'imaginary' vectorial potential. Such a model was introduced in1996 by Hatano and Nelson \cite{r6} in the context of flux lines in superconductors. They investigated the problem of Anderson localization in a disordered non-Hermitian lattice and showed that an imaginary magnetic field can prevent Anderson localization, with the appearance of a mobility interval at the center of the band (non-Hermitiain delocalization transition). Such a result was subsequently revisited by several authors \cite{r10} and connected to the problem of the spectrum of tridiagonal random matrices and random Dirac fermion models \cite{r11}. Recently, an optical implementation of the Hatano-Nelson model with an artificial maginary gauge field, based on a chain of coupled optical microrings with tailored gain and loss regions, was suggested \cite{r12} and the phenomenon of non-Hermitian transparency was disclosed \cite{r13}. In such previous studies \cite{r6,r10,r11,r12,r13,r13bis} the imaginary gauge field was considered stationary. However, it is well known that in ordinary tight-binding Hermitian quantum models oscillating electric and/or magnetic fields can deeply change the hopping dynamics via Peierls' substitution with important applications to coherent quantum state storage, dynamic decoupling and decoherence control (see, for instance, \cite{r14} and references therein). In one-dimensional lattices, oscillating fields renormalize the effective hopping rates and can result in coherent  destruction of tunneling and dynamic localization \cite{r15,r16}, which have been observed in matter wave and optical systems \cite{r17,r18}. In two-dimensional lattices, gauge fields are responsible for many important phenomena related to quantum Hall physics, topological  insulators and new phases of matter \cite{r18bis}. 
 
 \par
In this paper we present a theoretical study of the quantum dynamics of a particle hopping on a tight-binding lattice with a time-dependent (oscillating) imaginary gauge field, and highlight the role of topology on the structure of the quasi energy spectrum. For a quantum particle hopping on a ring threaded by an imaginary gauge flux, the energy spectrum in a stationary gauge field is always complex, however the addition of an ac (oscillating) gauge field can result in an entirely real quasi energy spectrum, i.e. the oscillating field can lead to a stabilization effect and pseudo-Hermitian dynamics (Sec.II). A fully different scenario is found for a particle hopping on a finite linear chain with open boundary conditions (Sec.III). In this case the energy spectrum is entirely real for a stationary imaginary gauge field, since the non-Hermitian problem with open boundary conditions is pseudo-Hermitian and can be mapped into an equivalent Hermitian model via an 'imaginary' gauge transformation. However, application of an oscillating imaginary gauge field breaks pseudo-Hermiticity and the quasi energy spectrum can become complex via multiple parametric resonances. A simple physical explanation of the different dynamical scenario  found in the two tight-binding lattices with different topology is also presented (Sec.IV).  

\section{Pseudo-Hermitian dyanmics in a tight-binding ring threaded by an oscillating imaginary gauge field}
Let us consider the hopping motion of a quantum particle on a tight-binding ring comprising $N \geq 3$ sites threaded by an imaginary and time-dependent gauge field $h=h(t)$; Fig.1(a). The non-Hermitian tight-binding Hamiltonian of the ring reads
\begin{equation}
\hat{H}(t)= \kappa \sum_{n=0}^{N-1} \left[ \exp(h) | n \rangle \langle n+1|+ \exp(-h) |n+1 \rangle \langle n | \right]
 \end{equation}
where $\kappa$ is the hopping rate, $h=h(t)$ is the imaginary gauge field, and the periodic ( Born-von Karman) boundary condition $|n+N \rangle=|n \rangle$ applies. After setting $|\psi(t) \rangle=\sum_{n=0}^{N-1} c_n(t) | n \rangle$, from the Schr\"{o}dinger equation $ i \partial_t |\psi(t) \rangle= \hat{H}(t) | \psi(t) \rangle$ the following coupled differential equations for the site amplitude probabilities $c_n(t)$ are found
\begin{equation}
i \frac{dc_n}{dt}= \kappa \exp[h(t)] c_{n+1}+\kappa \exp[-h(t)] c_{n-1}
\end{equation}
($n=0,1,2,...,N-1$) with the periodic boundary conditions
\begin{equation}
c_n(t)=c_{n+N}(t).
\end{equation} 
Let us first recall the properties of the energy spectrum of $\hat{H}$ in the stationary case $h(t)=h_0$ constant, which were discussed in previous works \cite{r6,r10}. The eigenfunstions and corresponding energies of $\hat{H}$ can be found from Eq.(2) by making the Ansatz 
\begin{equation}
c_{n}^{(l)}(t)=\exp(iq_l n-i E_l t)
\end{equation}
 where the wave number $q_l$ is quantized according to 
 \begin{equation}
 q_l= \frac{2 l \pi}{N}
 \end{equation} ($l=0,1,2,...,N-1$) because of the periodic boundary conditions (3).  Substitution of Eq.(4) into Eq.(2) yields
 \begin{equation}
 E_l=2 \kappa \cosh (h_0) \cos (q_l)+2 i \kappa \sinh (h_0) \sin(q_l).
 \end{equation}
Note that the energy spectrum is complex and the eigenvalues $E_l$ lie on the ellipse
\begin{equation}
\left[\frac{{\rm Re}(E)}{\cosh (h_0)} \right]^2+\left[\frac{{\rm Im}(E)}{\sinh (h_0)} \right]^2=4 \kappa^2,
\end{equation}
as already shown in previous works \cite{r6,r10}. An example of the energy spectrum is shown in Fig.1(b).\\
\begin{figure}[htbp]
  \includegraphics[width=84mm]{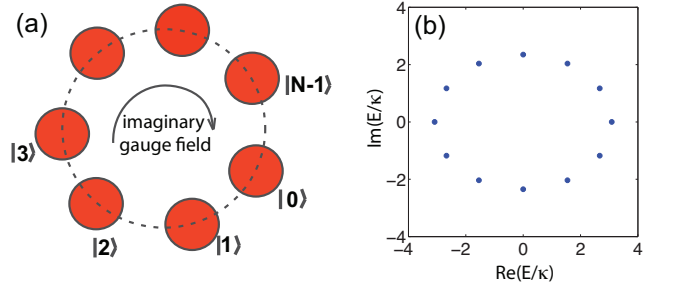}\\
   \caption{(color online) (a) Schematic of a tight-binding ring threaded by an imaginary gauge field $h=h(t)$. (b) Energy spectrum of the Hamiltonian (1) for a ring comprising $N=12$ sites in a stationary gauge field $h=h_0=1$. The energies in the complex plane lie on the ellipse defined by Eq.(7).}
 \end{figure}
The most general solution to Eq.(2) with $h(t)=h_0$ constant is given by an arbitrary superposition of eigenstates (4), i.e. $c_n(t)=\sum_{l=1}^N B_l \exp(iq_l n-i E_l t)$, where the amplitudes $B_l$ are determined by the initial condition $c_l(0)$. Taking into account that $\sum_{n=0}^{N-1} \exp(iq_l n-q_{\sigma} n)=N \delta_{l, \sigma}$, one readily obtains $B_{l}=(1/N) \sum_{n=0}^{N-1} \exp(-iq_l n) c_n(0)$, and thus
\begin{equation}
c_n(t)=\sum_{m=0}^{N-1} \mathcal{U}^0_{n,m}(h_0;t) c_m(0)
\end{equation}
where the propagator $\mathcal{U}^0_{n,m}(h_0; t)$ is defined by
\begin{equation}
\mathcal{U}^0_{n,m}(h_0; t) \equiv \frac{1}{N} \sum_{\sigma=0}^{N-1} \exp[ iq_{\sigma}(n-m)-iE_{\sigma} t]
\end{equation}
and $q_{\sigma}$, $E_{\sigma}$ are given by Eqs.(5) and (6). Note that in the Hermitian limit $h_0=0$ the ellipse defined by Eq.(7) shrinks into a line on the real axis (real energy spectrum) and the propagation $\mathcal{U}^{0}_{n,m}$ defined by Eq.(9) with $h_0=0$ is unitary and describes a quasi-periodic dynamics on the ring. \par
Let us now consider the more general case of a time-dependent imaginary gauge field $h=h(t)$. The most general solution to Eq.(2) is given by
\begin{equation}
c_n(t)=\sum_{m=0}^{N-1} \mathcal{U}_{n,m}(t) c_{m}(0)
\end{equation}
 where the propagator $\mathcal{U}_{n,m}(t)$ can be formally written as the ordered product
 \begin{eqnarray}
 \mathcal{U}(t) & = & \lim_{S   \rightarrow \infty} \prod_{k=1}^{S} \mathcal{U} ^0 (h_k; \Delta t) = \lim_{S \rightarrow \infty} \mathcal{U} ^0 (h_S; \Delta t) \times \nonumber \\
 &   \times & \mathcal{U} ^0 (h_{S-1}; \Delta t) \times .... \times \mathcal{U} ^0 (h_1; \Delta t)
 \end{eqnarray}
 where $\Delta t=t/S$, $t_k=k \Delta t$ and $\mathcal{U}^0(h; \Delta t)$ is defined by Eq.(9). For the propagator $\mathcal{U}^0$ defined by Eq.(9), the ordered product in Eq.(11) can be calculated in a closed form after some simple algebra. One obtains
 \begin{equation}
 \mathcal{U}_{n,m}(t)= \frac{1}{N} \sum_{\sigma=0}^{N-1} \exp \left[ iq_{\sigma}(n-m)-i \int_0^t dt' E_{\sigma}(t') \right]
 \end{equation}
where $E_{\sigma}(t)$ is defined by 
\begin{equation}
E_{\sigma} (t)=2 \kappa \cos (q_{\sigma}) \cosh [h(t)]+2i \kappa \sin (q_{\sigma}) \sinh [h(t)].
\end{equation}
Note that the propagator $\mathcal{U}(t)$ in the time-dependent case [Eq.(12)] is a simple extension of the propagator $\mathcal{U}^{0}(t)$ in the time-independent case [Eq.(9)] via the substitution $E_\sigma t \rightarrow \int_0^t dt' E_{\sigma} (t')$. Such a property mainly stems from the fact that the eigenfunctions of $\hat{H}$ with $h(t)=h_0$ constant are independent of $h_0$. Let us specialize our general result to the case of a time-periodic gauge field $h(t)$ with frequency $\omega$ and period $T=2 \pi / \omega$, i.e. $h(t+T)=h(t)$. In this case from Eq.(12) it readily follows that the $N$ quasi energies $E_l$ of the time-periodic Hamiltonian $\hat{H}(t)$ are given by
\begin{eqnarray}
E_l & = & 2 \kappa \cos q_l \left( \frac{1}{T} \int_0^T dt \cosh [h(t)]  \right) \nonumber \\
& + & 2 i \kappa \sin q_l \left( \frac{1}{T} \int_0^T dt \sinh [h(t)]  \right).
\end{eqnarray}
Like in the time-independent case, the quasi energies are complex and lie on the ellipse of equation 
\begin{equation}
\left[\frac{{\rm Re}(E)}{\frac{1}{T} \int_0^T dt \cosh h(t) } \right]^2+\left[\frac{{\rm Im}(E)}{\frac{1}{T} \int_0^T dt \sinh h(t) } \right]^2=4 \kappa^2,
\end{equation}
Interestingly, whenever the condition
\begin{equation}
 \int_0^T dt \sinh [h(t)] =0
\end{equation}
is satisfied, the quasi energy spectrum becomes real, and at the stroboscopic propagation times $t=T,2T,3T,...$ the time-periodic non-Hermitian Hamiltonian $\hat{H}(t)$ is equivalent to the effective stationary and Hermitian Hamiltonian of a tight-binding ring
\begin{equation}
\hat{H}_{eff}= \kappa_{eff} \sum_{n=0}^{N-1} \left( | n \rangle \langle n+1|+  |n+1 \rangle \langle n | \right]
\end{equation}
with enhanced hopping rate
\begin{equation}
\kappa_{eff}= \kappa \left( \frac{1}{T} \int_0^T dt \cosh [h(t)] \right).
\end{equation}
Therefore, provided that the condition (16) is met, the quantum dynamics of a particle on a ring described by the time-periodic non-Hermitian Hamiltonian (1) is pseudo-Hermitian and the role of the oscillating field is to stabilize the dynamics. Note that the condition (16) is satisfied for any arbitrary ac field $h(t)$ with zero mean satisfying the odd-symmetry constraint $h(-t+t_0)=h(t_0+t)$ for some $t_0$, for example for a sinusoidal field $h(t)=h_1 \sin (\omega t)$ regardless of the amplitude and frequency of the oscillation. Interestingly, the condition (16) can be met even for time-periodic fields $h(t)$ with a non vanishing dc term. For example, let us consider the piecewise constant field $h(t)$ defined by
\begin{equation}
h(t) \left\{
\begin{array}{cc}
h_1 & 0<t<T_1 \\
-h_2 & T_1 < t<T
\end{array}
\right.
\end{equation}
in the period $(0,T)$, with $h_{1,2}>0$. Provided that $h_1 T_1 \neq h_2 T_2$, where $T_2=T-T_1$, the field $h(t)$ has a non vanishing dc term. Further, if the condition $T_1 \sinh h_1=T_2 \sinh h_2$ holds, Eq.(16) is met and the dynamics is pseudo-Hermitian. Therefore the addition of an ac field to a dc imaginary gauge field can lead to stabilization (real quasi energy spectrum) and pseudo-Hermitian dynamics.\\
An example of pseudo-Hermitian dynamics on a ring comprising $N=3$ sites with a sinusoidal imaginary gauge field $h(t)=h_1 \sin (\omega t)$ is shown in Fig.2 for parameter values $\omega / \kappa=1$ and $h_1 / \kappa=0.4$. The solid curves in the figure depict the numerically-computed evolution of the site occupation amplitudes $|c_n(t)|$ at the three sites $n=0,1,2$ with the initial condition $c_n(0)=\delta_{n,0}$, corresponding to excitation of site $n=0$.  According to Eqs.(14) and (18), the quasi energies are  given by $E_l= 2 \kappa_{eff} \cos (2 \pi l /3)$ ($l=0,1,2$) with $\kappa_{eff}= \kappa (1/T) \int_0^T dt \cosh[h(t)] \simeq 2.081 \kappa$. i.e. there are two distinct quasi energies $E_{1,2}/ \kappa \simeq -1.0405, 2.081$. Since $|E_2-E_1|$ is incommensurate with the modulation frequency $\omega$, the dynamics turns out to be aperiodic. The dashed curves in the figure show the numerically-computed evolution of the occupation amplitudes at the three sites  for the effective stationary Hermitian Hamiltonian (17), which is periodic with period $\tau$ given by $\tau/T=\omega /|E_2-E_1| \simeq 0.32$. Note that, according to the theoretical analysis, at discretized times $t/T=1,2,3,....$ (vertical dotted curves in Fig.2) the dynamical behavior of the two Hamiltonians (1) and (17) does coincide.  

\begin{figure}[htbp]
  \includegraphics[width=84mm]{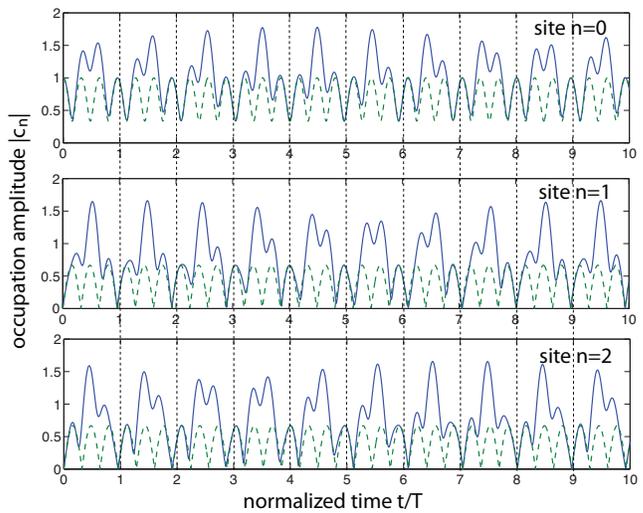}\\
   \caption{(color online) Pseudo-Hermitian dynamics in a ring, comprising $N=3$ sites, threaded by an imaginary gauge field $h(t)=h_1 \sin (\omega t)$ for $\omega / \kappa=1$ and $h_1/ \kappa=0.4$. The solid curves show the evolution of the site occupation amplitudes $|c_n(t)|$ at the three sites for the initial condition $c_n(0)=\delta_{n,0}$. The dashed curves show the corresponding behavior as obtained using the effective Hermitian Hamiltonian (17). The solid and dashed curves intersect at the discretized times $t/T=1,2,3,...$ (vertical dotted curves).}
 \end{figure}

\section{Multiple parametric resonance in a tight-binding linear chain with an oscillating imaginary gauge field}
Let us now consider the hopping motion of a quantum particle on a linear tight-binding chain comprising $N \geq 2$ sites with an imaginary and time-dependent gauge field $h=h(t)$; Fig.3. The non-Hermitian tight-binding Hamiltonian of the lattice reads
\begin{figure}[htbp]
  \includegraphics[width=80mm]{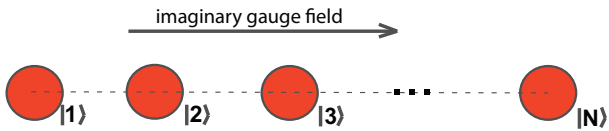}\\
   \caption{(color online) Schematic of a linear chain, comprising $N$ lattice sites with an imaginary and time-dependent gauge field. Left/right hopping rates are $\kappa \exp[\pm h(t)]$.}
 \end{figure}
\begin{equation}
\hat{H}(t)= \kappa \sum_{n=1}^{N-1} \left[ \exp(h) | n \rangle \langle n+1|+ \exp(-h) |n+1 \rangle \langle n | \right]
 \end{equation}
where $\kappa$ is the hopping rate and $h=h(t)$ is the imaginary gauge field. After setting $|\psi(t) \rangle=\sum_{n=1}^{N} c_n(t) | n \rangle$, from the Schr\"{o}dinger equation $ i \partial_t |\psi(t) \rangle= \hat{H}(t) | \psi(t) \rangle$ the following coupled differential equations for the site amplitude probabilities $c_n(t)$ are found
\begin{equation}
i \frac{dc_n}{dt}= \kappa \exp[h(t)] c_{n+1}+\kappa \exp[-h(t)] c_{n-1}
\end{equation}
($n=1,2,...,N$) with the open boundary conditions
\begin{equation}
c_0(t)=c_{N+1}(t)=0.
\end{equation} 
For a time-independent $h(t)=h_0$ imaginary gauge field, it is known that the energy spectrum of the non-Hermitian Hamiltonian (20) is entirely real and given by 
\begin{equation}
E_l=2 \kappa \cos (q_l),
\end{equation} 
where
\begin{equation}
q_l=\frac{l \pi}{N+1}
\end{equation}
$(l=1,2,...,N$). In fact, after the 'imaginary' gauge transformation $c_n(n)=a_n(t) \exp(-h_0n)$ the coupled-equations (21) yield
\begin{equation}
i \frac{da_n}{dt}= \kappa a_{n+1}+\kappa  a_{n-1}
\end{equation}
which describe the dynamics in an Hermitian tight-binding linear chain [Eq.(20) with $h=0$]. The $N$ eigenfunctions and corresponding energies of Eq.(25) are well known and given by $a^{(l)}_n(t)=(2/ \sqrt{N+1}) \sin (q_l n) \exp(-iE_l t)$, where $E_l$ and $q_l$ are defined by Eqs.(23) and (24). Therefore, in the stationary case the dynamics is pseudo-Hermitian. Using the above-mentioned gauge transformation, the most general solution to Eq.(21) with $h(t)=h_0$ can be readily found and reads  
\begin{equation}
c_n(t)=\sum_{l=1}^{N} \mathcal{U}^{0}_{n,l}(h_0; t) c_l(0)
\end{equation}
where the propagator $\mathcal{U}^0(h_0;t)$ is given by
\begin{eqnarray}
 \mathcal{U}^{0}_{n,l}(h_0; t)& = & \frac{2}{N+1} \exp[h_0(l-n)] \sum_{\sigma=1}^{N} \sin \left( \frac{\pi n \sigma}{N+1} \right) \;\;\; \\
 & \times &  \sin \left( \frac{\pi l \sigma}{N+1} \right) \exp \left[ -2 i \kappa t \cos \left( \frac{ \pi \sigma }{N+1} \right) \right]. \nonumber 
\end{eqnarray}
In the time-dependent case $h=h(t)$, according to Eq.(11) the propagator $\mathcal{U}(t)$ from $t=0$ to $t=t$ can be formally written as the ordered product of operators $\mathcal{U}^0(h_k; \Delta)$  of the stationary system, where $h_k=h(t_k)$ and $t_k=k \Delta t$. Unlike the ring lattice model considered in the previous section, in this case the product of operators cannot be determined in an analytical form and one has to resort to a numerical analysis. For a time-periodic gauge field with period $T= 2 \pi/ \omega$, i.e. $h(t+T)=h(t)$, according to Floquet theory the $N$ quasi energies of the time-periodic Hamiltonian (20)  are given by $E_l=(i/T) \rm{ln} ( \mu_l)$, where $\mu_l$ ($l=1,2,...,N$) are the $N$ eigenvalues of the matrix $\mathcal{U}(T)$, i.e. of the propagator over one oscillation period $T$. A numerical computation of quasi energies for an oscillating field shows rather generally that pseudo Hermitian dynamics can be broken and the quasi energy spectrum can become complex owing to the appearance of resonance tongues, which are the signature of a multiple parametric instability \cite{r20,r21,r22}. As an example, Fig.4 shows the numerically-computed instability domains (regions of complex quasi energies) in the frequency-amplitude plane $(\omega,h_1)$ for a square-wave ac gauge field 
\begin{equation}
h(t)= \left\{
\begin{array}{cc}
h_1 & 0<t<T/2 \\
-h_1 & T/2 < t<T
\end{array}
\right.
\end{equation}
and for a few increasing values of lattice sites $N$. For a square-wave modulation, the propagator $\mathcal{U}(T)$ over one oscillation cycle is readily computed as $\mathcal{U}(T)=\mathcal{U}^0(-h_1;T/2) \times \mathcal{U}^0(h_1;T/2)$, where $\mathcal{U}^0$ is defined by Eq.(27). The figure clearly indicates the appearance of resonance tongues emanating from $h_1=0$ at certain modulation frequencies $\omega$. The number of resonances increases as the number of site $N$ increases and become densely spaced close to $\omega \rightarrow 0$. In the simplest case of two sites ($N=2$), the instability arises from an ordinary parametric resonance, which for a non-Hermitian $\mathcal{PT}$ symmetric dimer has been recently studied in Ref.\cite{r23}. For a larger number of lattice sites in the chain, instability arises from multiple parametric resonances, which can be captured by a secular perturbation theory in the $h \rightarrow 0$ limit. Such an analysis is developed in the Appendix A. Rather generally, indicating by $l \omega$ the $l-th$ harmonic of the modulation function $h(t)$, resonance tongues emanate at the frequencies $\omega$ satisfying the condition
\begin{equation}
E_n-E_m \pm l \omega \simeq 0
\end{equation}
for some integers $n,m=1,2,...,N$, where $E_n= 2 \kappa \cos [n \pi /(N+1)]$.  For symmetry reasons, some of such resonances can be missed, as discussed in the Appendix for the simple case of $N=3$ lattice sites [Fig.4(a)]. As the number $N$ of lattice sites increases, the resonance conditions (29) are satisfied in a densely number of frequencies below the cut-off frequency $4 \kappa$, as shown e.g. in Fig.4(d) for $N=50$ sites.
Since $E_n$ falls inside the range $(-2 \kappa, 2 \kappa)$, for a modulation frequency $\omega$ larger than $4 \kappa$, the resonance condition (29) can never be satisfied. Therefore, for $\omega > 4 \kappa$ parametric resonances are prevented and the quasi energies are entirely real (for a not-to-large value of the modulation amplitude). Such a result is in agreement with the fact that, in the large modulation frequency limit, one can average the rapidly-oscillating imaginary Peierls's phases in Eq.(21), leading to an effective Hermitian linear chain with hopping rate $(\kappa /T) \int_0^T dt \exp[\pm h(t)]$ (see, for instance, \cite{r24}).\\ 
\begin{figure}[htbp]
  \includegraphics[width=89mm]{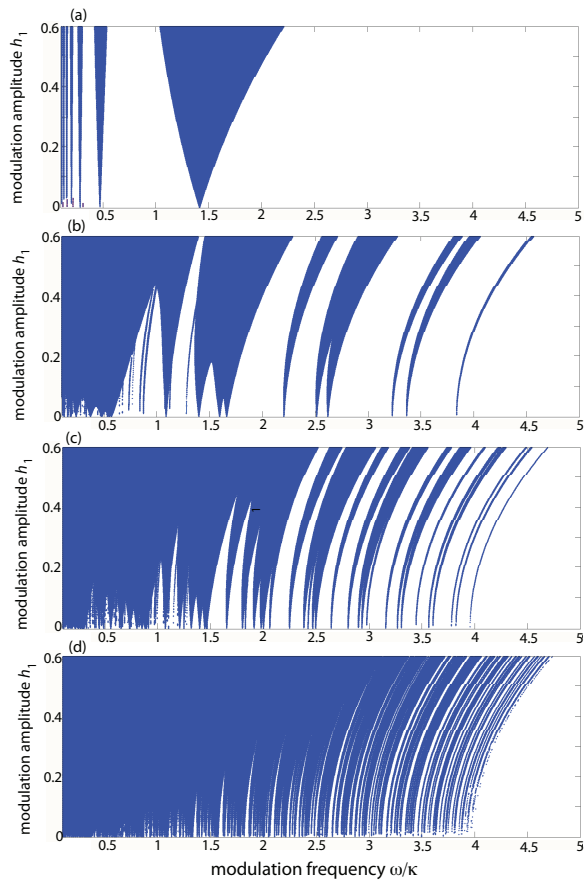}\\
   \caption{(color online) Numerically-computed regions of complex quasi energies (dark areas) arising from multiple parametric resonances for a square-wave modulation $h(t)$ and  for an increasing number $N$ of lattice sites: (a) $N=3$, (b) $N=10$, (c) $N=20$, and (d) $N=50$. The number of resonance tongues rapidly increases, below the cut-off frequency $ 4 \kappa$, as $N$ increases.}
 \end{figure}
An example of parametric instability  on a linear chain comprising $N=3$ sites with a sinusoidal imaginary gauge field $h(t)=h_1 \sin (\omega t)$ is shown in Fig.5 for parameter values $\omega / \kappa=1$ and $h_1 =0.4$ in Fig.5(a), and $\omega / \kappa= \sqrt{2}$ and $h_1 =0.4$ in Fig.5(b). The curves in the figure depict the numerically-computed evolution of the site occupation amplitudes $|c_n(t)|$ at the three sites $n=1,2,3$ with the initial condition $c_n(0)=\delta_{n,1}$, corresponding to excitation of the edge site $n=1$. While in Fig.5(a) the modulation frequency is far from any resonance tongue and the dynamics is pseudo-Hermitian (real quasi energies), in Fig.5(b) the modulation frequency is set in resonance with the first resonance tongue [see Fig.4(a)] and an instability is clearly observed, corresponding to a secular growth of amplitudes $|c_n(t)|$ and complex quasi energy spectrum. 

\begin{figure}[htbp]
  \includegraphics[width=84mm]{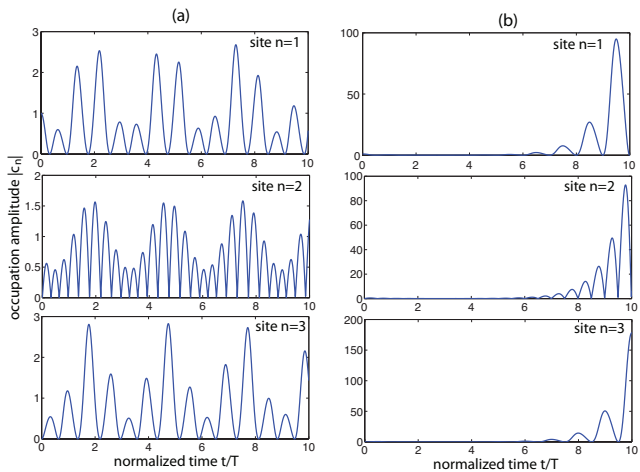}\\
   \caption{(color online) Parametric instability in a linear chain comprising $N=3$ sites with an oscillating imaginary gauge field $h(t)=h_1 \sin (\omega t)$ for (a) $\omega / \kappa=1$, $h_1=0.4$, and (b) $\omega / \kappa= \sqrt{2}$, $h_1=0.4$. The curves in the various panels show the evolution of the site occupation amplitudes $|c_n(t)|$ at the three sites for the initial condition $c_n(0)=\delta_{n,1}$. In (a) the quasi energy spectrum is real and the dynamics pseudo-Hermitian. In (b) the quasi energy spectrum becomes complex owing to parametric resonance, which is clearly manifested in the secular growth of amplitudes $|c_n(t)|$.}
 \end{figure}

\section{Pseudo-Hermitian dynamics and parametric instability: a simple physical description}
In the previous two sections we have shown that, when an oscillating imaginary gauge field is applied to a tight-binding lattice, two different phenomena can arise depending on the lattice topology, i.e.on boundary conditions: pseudo-Hermitian dynamics in a ring lattice, and parametric resonances in a linear chain. In this section we show that such phenomena can be explained in a simple physical way by considering the limit of a large number of sites $N$ in the lattice. In such a limit, the two phenomena can be captured by considering a wave packet that either travels along a ring or propagates back and forth in a linear chain. Let us consider an infinitely-extended lattice in the presence of a stationary imaginary gauge field $h$. As shown in Ref.\cite{r12},  wave transport in the lattice is highly asymmetric because wave packets propagating in opposite directions are one amplified and the other one damped. Such a property follows from the nature of the dispersion relation of plane waves $c_n(t) \sim \exp[iqn-iE(q)t]$ in the infinitely extended lattice, which reads \cite{r12}
\begin{equation}
E(q)=2 \kappa \cosh(h) \cos q +2i \kappa \sinh(h) \sin q,
\end{equation}    
where $-\pi<q<\pi$ is the Bloch wave number. A wave packet, obtained by a superposition of plane waves around a carrier wave number $q$, propagates with a group velocity $v_g=(d {\rm Re}(E) /dq)=-2 \kappa \cosh h \sin q$. For $h>0$, a forward propagating wave packet ($-\pi<q< 0$, $v_g>0$) is attenuated since ${\rm Im}(E)<0$, whereas a backward propagating wave packet ($0<q< \pi$, $v_g<0$) is amplified since ${\rm Im}(E)>0$. The opposite behavior occurs when the sign of the gauge field $h$ is reversed; see Fig.6. \par
\begin{figure}[htbp]
  \includegraphics[width=70mm]{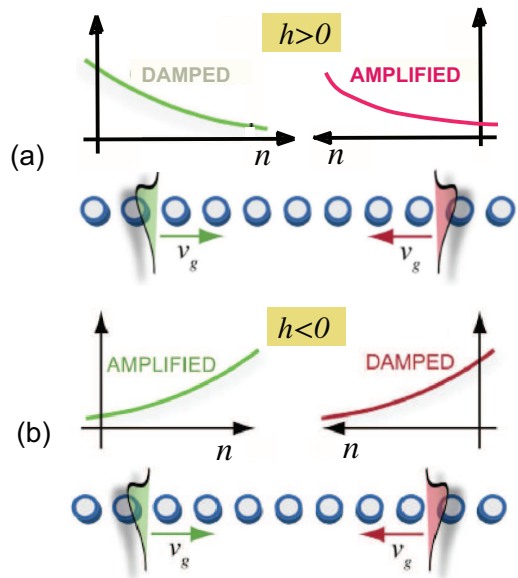}\\
   \caption{(color online) Schematic of wave packet propagation in the bulk of a tight-binding lattice with an imaginary gauge field $h$. For $h>0$ [panel (a)] a forward (backward) propagating wave packet is damped (amplified), whereas for $h<0$ [panel (b)] the reversed behavior occurs.}
 \end{figure}
 With such a property in mind, let us first consider propagation of a wave packet in a ring. The ring periodic boundary conditions just introduce a quantization of the wave number $q$, which however for a large number sites $N$ and sufficiently localized wave packet does not change the dynamics as compared to the infinitely extended lattice. Therefore, for a stationary field a localized wave packet is either damped or amplified secularly, depending on the circulation direction (either clockwise or counter-clockwise). This is in agreement with the fact that for a time-independent field $h$ the energy spectrum  of the Hamiltonian is complex. However, when the gauge field $h(t)$ oscillates in time, the wave packet undergoes periodic amplification and attenuation, regardless of the circulation direction. In particular, whenever the time-average of the amplification/attenuation rate ${\rm Im}(E)=2 \kappa \sin [h(t)] \sin q$ vanishes, secular growth or decay of the wave packet is suppressed, and the dynamics is pseudo-Hermitian. Note that the vanishing on average of the  amplification/attenuation rate is precisely expressed by the condition (16) derived in Sec.II.\\
 Let us now consider the propagation of a wave packet in a long chain.  For a stationary imaginary gauge field $h$, a forward propagating wave packet is attenuated, however when it reaches the right chain boundary it is reflected, and the backward-propagating wave packet is amplified at the same rate. At the left chain boundary, the wave packet is reflected and the forward-propagating wave packet is damped. Hence a periodic attenuation/amplification in a balanced manner occurs after each reflection at the left/right lattice edges: the dynamics is thus pseudo-Hermitian with no secular growth or attenuation of the wave packet amplitude on average. When the imaginary gauge field $h(t)$ is an oscillating field with zero mean, $h(t)$ changes sign within each oscillation cycle and the dynamics is strongly dependent on the ratio between the oscillation period and the transit time of the wave packet in the chain. Let us assume for the sake of definiteness that $h(t)$ is positive in the first half semi cycle of oscillation $0<t<T/2$ and negative in the second semi cycle $T/2<t<T$. This is the case, for example, of a square-wave or of a sinusoidal field. The transit time of the wave packet between the two edges of the chain is $\tau=N/v_g$. If the transit time in an odd multiple than $T/2$, i.e. for $\tau=(2l+1)T/2$, an overall amplification in one circulation direction occurs, which cumulates at successive transits back and forth between the two lattice edges. This is because after each reflection at the lattice edge the sign of $h$ changes synchronously. As a result, a secular growth of the wave packet amplitude arises in one circulation direction, which is the signature of the parametric instability. Taking into account that for $|h(t)| \ll1$ one has $\tau=N/v_g \simeq N/ (2 \kappa |\sin q|)$,  the condition $\tau=(2l+1)T/2=(2l+1)\pi / \omega$ yields $\omega=2 \kappa \pi (2l+1) |\sin q| /(N)$. For a long chain, the frequencies at which parametric instability arises are thus a dense set, according to the numerical results [see Fig.4(d)]. The existence of a cut off frequency, above which parametric instability is suppressed, follows from the requirement that the oscillation period $T$ must be longer than the characteristic reflection time $\tau_R$ of the wave packet, i.e. $T> \tau_R$.  For a wave packet with lattice extension $\Delta n$ and group velocity $v_g$, the reflection time can be estimated as $\tau_R \sim \Delta n / v_g$. 
For a narrow wave packet with lattice extension of a few sites, e.g. taking $\Delta n \sim  \pi$, the  shortest reflection time is obtained at the carrier Bloch wave number $q= \pi/2$, corresponding to the largest group velocity $v_g \simeq 2 \kappa$ and thus to a reflection time $\tau_R \sim \pi /  (2 \kappa)$. The requirement $T> \tau_R$ thus gives $\omega < \sim 4 \kappa$, which is precisely the cut-off condition rigorously derived in Sec.III. 
 
 \section{Conclusion}
 Driven tight-binding lattices provide a fertile quantum model for coherent quantum control, dynamic decoupling and decoherence control  in quantum physics \cite{r14,r15,r16}. The application of oscillating electric or magnetic fields on a particle hopping on a lattice introduces Peierls' phases that can be tailored to realize such important effects as hopping rate renormalization, coherent  destruction of tunneling, dynamic localization, and quantum Hall physics \cite{r15,r16,r17,r18}. While great attention has been devoted so far to study Peierls' phase in driven Hermitian systems and the related broad fields of artificial gauge fields and novel phases of matter, the effects of an oscillating {\it imaginary} gauge field  have been so far overlooked. Imaginary gauge fields were introduced in a pioneering paper by Hatano and Nelson \cite{r6} to study non-Hermitian Anderson localization in disordered lattices, which raised a lively interest \cite{r10,r11}. Recently, the proposal to implement artificial imaginary gauge fields in integrated photonics using coupled optical microrings with tailored gain and loss regions \cite{r12,r13} has renewed the interest in the Hatano-Nelson model, paving the way toward an experimental demonstration of non-Hermitian Anderson delocalization transition. Such previous studies, however, are limited to consider stationary imaginary gauge fields. In this work we theoretically investigated the quantum dynamics in a one-dimensional tight-binding lattice with an oscillating imaginary gauge field. As compared to the analogous problem of real gauge fields, the imaginary gauge field problem discloses a completely different dynamical behavior, which is strongly sensitive to lattice topology even in the one-dimensional case.  For a quantum particle hopping on a ring threaded by an imaginary gauge flux, the energy spectrum in a stationary gauge field is always complex, however the addition of an ac (oscillating) gauge field can result in an entirely real quasi energy spectrum and pseudo-Hermitian dynamics. Conversely, if the particle hops on a finite linear chain with open boundary conditions, the energy spectrum is entirely real for a stationary gauge field but can become complex via multiple parametric resonances in an oscillating field. Our results highlight the very different physics of  tight-binding lattices driven by either real or imaginary gauge fields, providing important novel insights into the dynamical behavior in the non-Hermitian case. The  present analysis could be extended into several directions, for example by considering mixed real and imaginary gauge fields as well as two-dimensional lattice geometries.

\appendix

\section{Multiple parametric resonances in a linear lattice with open boundary conditions: secular perturbation theory}
In this Appendix we present a secular perturbation theory of Eq.(21)  showing the appearance of multiple parametric resonances, leading to complex quasi energies, in the limit of a small gauge field $h(t) \rightarrow 0$. Without loss of generality, we assume that $h(t)$ is an ac field with zero mean; a non vanishing dc value $h_0$ of the periodic function $h(t)$ can be  eliminated by the gauge transformation $c_n(n) \rightarrow c_n(t) \exp(-h_0n)$ and hence it does not affect the quasi energy spectrum of the Hamiltonian (20). For $|h(t)| \ll 1$, one can write $\exp[ \pm h(t)] \simeq 1 \pm h(t)$ and Eq.(21) take the form
\begin{equation}
i \frac{d \mathbf{c}}{dt}= \mathcal{A} \mathbf{c}+h(t) (\mathcal{B}_1-\mathcal{B}_2) \mathbf{c},
\end{equation}
where $\mathbf{c} \equiv (c_1,c_2,...,c_N)^T$ is the vector of the site occupation amplitudes and the $N \times N$ matrices $\mathcal{A}, \mathcal{B}_1$ and $\mathcal{B}_2$ are defined by
\begin{eqnarray}
\mathcal{A}_{n,m} & = & \kappa (\delta_{n,m-1}+\delta_{n,m-1}) \\
\left( \mathcal{B}_{1} \right)_ {n,m} & = & \kappa \delta_{n,m-1} \\
\left( \mathcal{B}_{2} \right)_ {n,m} & = & \kappa \delta_{n,m+1} \\
\end{eqnarray}
($n,m=1,2,...,N$). Let us indicate by $\mathcal{T}$ and $\mathcal{E}$ the eigenvector matrix and corresponding diagonal eigenvalue matrix of $\mathcal{A}$, i.e. such that $\mathcal{A} \mathcal{T}= \mathcal{T} \mathcal{E}$. The explicit forms of $\mathcal{T}$ and $\mathcal{E}$ read
\begin{eqnarray}
\mathcal{T}_{n,m} & = & \sqrt{\frac{2}{N+1}} \sin \left( \frac{n m \pi}{N+1} \right) \\
\mathcal{E}_{n,m} & = & E_n \delta_{n,m} = 2 \kappa \cos \left( \frac{n \pi}{N+1} \right) \delta_{n,m}
\end{eqnarray} 
After setting $\mathbf{c}(t)= \mathcal{T} \mathbf{a}(t)$, i.e. in the basis that diagonalizes $\mathcal{A}$, Eq.(A1) can be cast in the form
\begin{equation}
i \frac{d \mathbf{a}}{dt}= \mathcal{E} \mathbf{a}+h(t) \mathcal{P} \mathbf{a},
\end{equation}
where we have set
\begin{equation}
\mathcal{P} \equiv \mathcal{T}^{-1}(\mathcal{B}_1-\mathcal{B}_2) \mathcal{T}.
\end{equation}
Taking into account that $\mathcal{T}^{-1}= \mathcal{T}$, after some cumbersome algebra one can derive the following expression of the elements of the time-independent perturbation matrix $\mathcal{P}$ entering in Eq.(A8)
\begin{eqnarray}
\mathcal{P}_{n,m} & = & \kappa \frac{1-(-1)^{n+m}}{N+1} \sin \left( \frac{m \pi}{N+1} \right) \\
& \times & \left[ {\rm{cotg}} \frac{ \pi(n+m)}{2(N+1)} +{\rm{cotg}} \frac{\pi(n-m)}{2(N+1)} \right]. \nonumber 
\end{eqnarray}
Note that $\mathcal{P}$ is an anti-Hermitian matrix, i.e. $\mathcal{P}_{n,m}=-\mathcal{P}_{m,n}^{*}$, and $\mathcal{P}_{n,m}$ vanishes when $|n+m|$ is an even number. In the absence of the oscillating gauge field $h=0$, the dynamical system described by Eq.(A8) is neutrally stable since the energies $E_n$ are real. The addition of the perturbation term on the right hand side of Eq.(A8) can lead to secularly growing terms via typical parametric resonance phenomena, corresponding to complex quasi energies. To capture the onset of parametric resonances, we perform a rather standard secular perturbation analysis of Eq.(A8) by letting $h(t) \rightarrow \alpha h(t)$, where $\alpha$ is a smallness parameter that indicates the order of magnitude of the various terms entering in the asymptotic analysis (see, for instance, \cite{r20,r22}). We look for a solution to Eq.(A8) as a power series in $\alpha$
\begin{equation}
\mathbf{a}= \mathbf{a}^{(0)}+ \alpha \mathbf{a}^{(1)}+ \alpha^2 \mathbf{a}^{(2)}+...
\end{equation}
and introduce multiple time scales
\begin{equation}
T_0=t \:, \;\; T_1= \alpha t \; , \;\; T_2= \alpha^2 t \;, ...
\end{equation}
which are necessary to remove secular growing terms that would prevent the asymptotic expansion (A11) to be uniformly valid in time. Substitution of Eq.(A11) into Eq.(A8) and using the derivative rule $(d/dt)=(d/dT_0) + \alpha (d/dT_2)+ \alpha^2 (d/dT_2)+...$ yields a hierarchy of equations at successive orders in $\alpha$. At lowest order $\sim \alpha^ 0$ one obtains $i ( da_{n}^{(0)} /dT_0)=E_n a_n^{(0)}$, which yield
\begin{equation}
a_n^{(0)}=A_n(T_1,T_2,...) \exp(-i E_n T_0).
\end{equation}
 where the amplitudes $A_n$ are allowed to vary on the slow time scales $T_1,T_2,...$. At order $ \sim \alpha$ one obtains
 \begin{equation}
 \left(i  \frac{d}{dT_0}-E_n \right) a_n^{(1)}=G_n^{(1)}(T_0)
 \end{equation}
 where we have set
 \begin{eqnarray}
 G_n^{(1)} & \equiv & - i \frac{\partial  A_n}{\partial T_1} \exp(-i E_n T_0) \\
 & + &  h(T_0) \sum_{m=1}^N \mathcal{P}_{n,m} A_m \exp(-i E_m T_0). \nonumber
 \end{eqnarray}
To avoid the appearance of secularly growing terms when solving Eq.(A14), the driving term $G_n^{(1)}$ defined by Eq.(A15) should not contain a term oscillating like $\sim \exp(-i E_n T_0)$. The solvability conditions thus yield the following coupled equations for the evolution of the amplitudes $A_n$ on the slow time scale $T_1$
\begin{equation}
i \frac{dA_n}{dT_1}= \sum_{m=1}^{N} \mathcal{R}_{n,m} \mathcal{A}_m,
\end{equation}
where we have set
\begin{equation}
\mathcal{R}_{n,m} \equiv \mathcal{P}_{n,m} \langle h(t) \exp [i(E_n-E_m)t] \rangle
\end{equation}
and the brackets $\langle .. \rangle$ on the right hand side of Eq.(A17) denotes time average of the oscillating term. Since $h(t)$ is real and periodic with period $T= 2 \pi/ \omega$, the matrix element $\mathcal{R}_{n,m}$ does not vanish provided that the resonance condition
\begin{equation}
E_n-E_m \pm l \omega \simeq 0
\end{equation}
 is satisfied for some integer $l$, with $ l \neq 0$ \cite{note}. This is the resonance condition given by Eq.(29) in the text. From Eqs.(A10) and (A17), it can be readily shown that the matrix $\mathcal{R}$ is anti-Hermitian, i.e. its eigenvalues are purely imaginary. Moreover, if $\lambda$ is an eigenvalue of $\mathcal{R}$, then $\lambda^{*}$ is an eigenvalue as well. Therefore, provided that some of the elements of the matrix $\mathcal{R}$ do not vanish, the solution to Eq.(A16) shows secularly growing terms on the time scale $T_1$, i.e. the quasi energies of the time-periodic system (A1) become complex with an imaginary part of order $\sim \alpha$. This explains the existence of multiple parametric resonance tongues found in the numerical computation of the quasi energies, shown in Fig.4. As the number of lattice sites $N$ increases, the number of resonance tongues, as determined by the resonance condition (A18), rapidly increases below the cut-off frequency $ 4 \kappa$.\\ 
 It should be noted that, since $\mathcal{R}_{n,m}$ vanishes when $|n-m|$ is an even number, some of the resonance tongues predicted by Eq.(A18)  can be missed. Let us discuss, for example, the simplest case of $N=3$ lattice sites. In this case the three eigenvalues $E_l$ read
 \begin{equation}
 E_1=\sqrt{2} \kappa \; , \; E_2=0 \; , \; E_3=-\sqrt{2} \kappa.
 \end{equation}
 Therefore, according to Eq.(A18) resonance tongues are expected to emanate from the two sets of frequencies
 \begin{eqnarray}
 \omega_l^{(1)} & = & \frac{|E_3-E_1|}{l} = \frac {2\sqrt{2} \kappa} {l} \\
 \omega_l^{(2)} & = & \frac{|E_2-E_1|}{l} = \frac{|E_3-E_2|}{l} =\frac {\sqrt{2} \kappa} {l} 
 \end{eqnarray}
 with $l=1,2,3,...$. However, since $\mathcal{P}_{13}=\mathcal{P}_{31}=0$, the family of resonances $\omega^{(1)}_l$ is missed. Moreover, for a square wave modulation like the one considered in Fig.4 only odd Fourier amplitudes of $h(t)$ do not vanish, i.e. for the family of resonances $\omega^{(2)}_l$  only those with an odd integer $l$ should be considered. Therefore, the actual resonance tongues in a linear chain with $N=3$ sites emanate from the frequencies $\sqrt{2} \kappa$, $\sqrt{2} \kappa/3$, $\sqrt{2} \kappa/5$, ..., in agreement with the numerical results shown in Fig.4(a).

%\clearpage
%\bibliography{H:/Physik/bibliography}

\begin{thebibliography}{31}
\bibitem{note}
For a sinusoidal modulation $h(t)=h_1 \sin (\omega t)$, only the first harmonic $l=1$ should be taken in Eq.(A18), since higher-order Fourier harmonic amplitudes vanish for a sinusoidal modulation. This holds in first-order perturbation theory, i.e. u to the time scale $T_1$, which captures first-order resonance tongues. However, pushing the asymptotic analysis to higher orders, i.e. by considering the evolution of the amplitudes $A_n$ at the longer time scales $T_2$, $T_3$,...,  higher harmonics of the modulation frequency $\omega$ appear even for a sinusoidal modulation. This is well known, for example, for the usual asymptotic theory of parametric resonance in the Mathieu equation. Hence Eq.(A18) with arbitrary $l$ can be regarded as the general multiple parametric resonance condition.
\end{thebibliography}

\end{document}